\documentclass[12pt]{iopart}

\usepackage{iopams}
\usepackage{color}
\usepackage{graphicx}
\usepackage{dcolumn}
\usepackage{bm}
\usepackage[utf8]{inputenc}
\usepackage[T1]{fontenc}
\usepackage{mathptmx}
\usepackage{theorem}
\usepackage{xcolor}
\usepackage{caption}
\usepackage{subcaption}
\usepackage{cite}

\newtheorem{thm}{Theorem}

\newtheorem{defi}{Definition}

\newcommand{\proof}{\noindent {\bf Proof:} \hspace{0.1in}}
\newcommand{\qed}{\hfill\mbox{\raggedright $\Box$}\medskip}

\newcommand{\R}{\mathbb{R}}

\newcommand{\M}{\mathsf{M}}
\newcommand{\MS}{\mathsf{S}}

\newcommand{\I}{\mathsf{I}}
\newcommand{\metric}{\ensuremath{\mathrm{g}}}
\newcommand{\submetric}{\ensuremath{\mathrm{h}}}
\newcommand{\submetrict}{\ensuremath{\gamma}}
\newcommand{\normal}{\ensuremath{\mathrm{u}}}

\newcommand{\grad}{\raisebox{0.01em}{\scalebox{1.1}[1.4]{${\scriptstyle \nabla}$}}}
\newcommand{\gradbar}{\raisebox{0.01em}{\scalebox{1.1}[1.4]{${\scriptstyle \overline{\nabla}}$}}}
\newcommand{\gradbarD}{\raisebox{0.01em}{\scalebox{1.2}[1.0]{${\mathrm{D}}$}}}
\newcommand{\energy}{\ensuremath{\mathrm{\rho}}}

\newcommand{\pressure}{\scalebox{0.9}[1.0]{\ensuremath{\mathsf{P}}}}
\newcommand{\anisotropy}{\scalebox{1.1}[1.0]{\ensuremath{\Sigma}}}

\newcommand{\curvature}{\ensuremath{\overline{\mathrm{R}}}}

\newcommand{\Lie}{\ensuremath{\mathfrak{L}}}
\newcommand{\hubble}{\ensuremath{\mathrm{H}}}

\newcommand{\hubblevec}{\ensuremath{\mathrm{X}_H}}
\newcommand{\shear}{\ensuremath{\sigma}}

\begin{document}

\title[Spacetimes with homogeneous and isotropic expansion]{Spacetimes with homogeneous and isotropic expansion}

\author{L. G. Gomes}
\address{Federal University of Itajub\'a (UNIFEI), Av. BPS, 1303, Itajub\'a-MG, 37500-903, Brazil}
\eads{\mailto{lggomes@unifei.edu.br}}

\date{\today}

\begin{abstract}
In this short note, we define and characterize all the spacetimes admitting observers to whom the cosmic expansion is homogeneous and isotropic and interpret their Einstein's equations.    
\end{abstract}

\maketitle

The standard model of cosmology is based on spacetimes where expansion is homogeneous and isotropic for a special class of free-falling observers to whom the proper time and the Hubble parameter do not change along spatial directions \cite{Book_2008_Weinberg}. Notwithstanding, today we know the universe cannot be considered homogeneous on scales less than 50 Mpc \cite{MNRAS_2023_Bernui}, and maybe even on larger distances \cite{ObservationalCosmologicalPrinciple}. Is it possible to have homogeneous and isotropic expansion living together with inhomogeneities? Which characteristics does such a hypothesis impose on spacetime? In this note, we address such questions.

We start by picking a set of cosmological observers $\normal$, which, following Ref. \cite{LGGomes_2024_CQG_1}, is a time-like, vorticity-free, oriented to the future, and unit vector field in the spacetime $\M$ with metric $\metric$. Any $\normal$-observer $c(\tau)$, formally an integral curve of $\normal$, perceives the spatial volume of the universe expanding at a rate\footnote{See Ref. \cite{LGGomes_2024_CQG_1} for the details of the concepts used here.}  $\hubble(c(\tau))$, its Hubble parameter. A (volumetric) scalar factor is any function $a(\tau)$ along this observer such that, in terms of its proper time $\tau$, it characterizes this expansion rate as   
\begin{equation}\label{Eq:HubbleGeneralCoordinates}
\hubble (c(\tau)) = \frac{1}{a}\frac{d a}{d \tau} \, . 
\end{equation}

When we think of the dynamics of our Universe, we very often assume the existence of a ubiquitous scale factor that would have the same properties despite the $\normal$-observer's location, demanding the spatial volume to expand equally for all of them. This encompasses a weak notion of the cosmological principle, the driving tenet of modern Cosmology. In this way, both the $\normal$-observers inside massive clusters of galaxies and those in the middle of the big cosmic voids should be able to perceive a ubiquitous scale factor, which does not change among them, that is, is spatially homogeneous. This would imply a natural and universal pace for cosmic evolution. 

Formally, this notion of homogeneity in cosmic evolution would be naively described by imposing the condition $\gradbar \hubble=0$, where $\gradbar$ stands for the projection of the covariant derivative $\nabla$ of spacetime metric $\metric$ along the $\normal$-spatial sections. However, in the $\normal$-frame of reference, a $\normal$-observer experiments a Newtonian-like acceleration, namely, $\gradbar\phi(c(\tau))=\grad_\normal\normal(c(\tau))$. As we define locally a time $t$ characterizing the $\normal$-spatial sections by $t=\textrm{const}$, the proper time $\tau$, which is also the parameter of the flux $F_\normal^\tau(p)$ of $\normal$, is $d\tau = e^{\phi(c(\tau))}dt$. In other words, the Hubble parameter varies along the $\normal$-observers due to the choice for the scale factor $a(c(\tau))$ and also to their different proper times. Hence, the condition $\gradbar \hubble=0$ is suitable for describing homogeneity of expansion only when all the observers are free-falling, that is, $\grad_\normal\normal=0$, as in the FLRW \cite{Book_2012_ellis_mac_marteens} and generalized RW \cite{IJGMMP_2017_GeneralizedRW_Survey} models. Therefore, we must withdraw the proper time effects from this condition for an adequate definition of homogeneous expansion. For this reason, we come up with the Hubble homogeneity vector field, which is set as
\begin{equation}
\hubblevec = \gradbar\hubble + \hubble \grad_\normal\normal \, .
\end{equation}

On the other hand, the notion of isotropy of expansion is standard: the trace-less part of the expansion tensor, the shear $\sigma_{\mu\nu}$ \cite{Book_2012_ellis_mac_marteens}, or better, the anisotropy tensor $\anisotropy_{\mu\nu} \sim \sigma_{\mu\nu}/\hubble$ \cite{LGGomes_2024_CQG_1, LGGomes_2022_IJMPD, LGGomes_2023_EPJC_1}, measures the departures from isotropy as they are perceived along by the $\normal$-observers. With those ideas in mind, we set:
\begin{defi}
The Hubble evolution (cosmological expansion/contraction) is homogeneous for the $\normal$-observers if $X_\hubble=0$. It is isotropic for the $\normal$-observers when it is shear-free, $\sigma_{\mu\nu}=0$. When both conditions are satisfied for the same $\normal$-observers, we say the Hubble evolution is homogeneous and isotropic. 
\end{defi}

In the following, we denote $\normal^\flat=\normal_\mu dx^\mu$, $\Lie_\normal$ the Lie derivative along $\normal$,  and $\MS_{p}$ the $\normal$-spatial section at $p$, the maximal integral manifold of the distribution $\normal^\bot$ passing at $p \in \M$. Any function $f$ which is homogeneous along the $\normal$-spatial sections, that is, $\gradbar f = 0$, we often denote by $f(t)$. Following these considerations, we have:
\begin{thm}\label{Thm:HubbleHomogeneityIsotropy}
Let $\normal$ be a unit, vorticity-free, and time-like vector field in a spacetime $(\M,\metric)$ such that none of the $\normal$-integral curves are closed. Denote the induced $\normal$-spatial metric by $\submetric = \metric + \normal^\flat\normal^\flat$ and, for any $p_0\in \M$, the maximal set obtained by carrying $\MS_{p_0}$ along the flux of $\normal$ by $U$. It follows that $U$ is open in $\M$ and there is a unique smooth function $a: U \to \R$ satisfying $a=1$ in $\MS_{p_0}$ such that, when restricted to any $\normal$-observer in $U$, it is also a scale factor for them. Moreover, as we take the $\normal$-spatial metric on $U$ set by $\submetrict=\submetric/a^2$ together with its natural volume form $\widehat{\Omega}(\submetrict)$, we have:
\begin{enumerate}
\item The Hubble evolution is homogeneous for the $\normal$-observers in $U$ if, and only if, the scale factor with $a=1$ in $\MS_{p_0}$ satisfies $\gradbar a=0$ with the volume form $\widehat{\Omega}(\submetrict)$ invariant by $\normal$. In short, $\hubblevec =0$ if, and only if, 
\begin{equation}\label{Eq:ThmHubbleHomogeneity}
\submetric = a(t)^2\, \submetrict  \quad \textrm{with} 
\quad \Lie_\normal\widehat{\Omega}(\submetrict) =0 \, .
\end{equation}
\item The Hubble evolution is isotropic for the  $\normal$-observers in $U$ if, and only if, the spatial metric $\submetrict$ is invariant by the flux of $\normal$. In short, $\shear_{\mu\nu} =0$ if, and only if,
\begin{equation}\label{Eq:ThmHubbleIsotropy}
\quad \Lie_\normal\submetrict =0 \, .
\end{equation}
%
%
\item The Hubble evolution is homogeneous and isotropic for the  $\normal$-observers if, and only if, the scale factor $a: U \to \R$ with $a=1$ in $\MS_{p_0}$ satisfies $\gradbar a=0$ and the spatial metric $\submetrict$ is invariant by $\normal$. In short, $\sigma_{\mu\nu}=0$ and $\hubblevec =0$ if, and only if,
\begin{equation}\label{Eq:ThmHubbleHomogeneityIsotropy}
\submetric = a(t)^2\, \submetrict  \quad \textrm{with} \quad \Lie_\normal\submetrict =0 \, .
\end{equation}
\end{enumerate}
\end{thm}
\proof
Denoting the flux of $\normal$ by $F^{\tau}_\normal$ and fixing $p_0 \in \M$, the (volumetric) scale factor is uniquely determined up to a multiplicative positive function $a_0: \MS_{p_0} \to \R$ such that, for each $p \in \MS_{p_0}$,
\begin{equation}\label{Eq:ScaleFactorvsFluxNormal}
a\left(F^{\tau}_\normal(p)\right) = a_0(p) \mathrm{exp}\left( \int_0^\tau \hubble \left(F^{s}_\normal(p)\right)\, ds \right) \, . 
\end{equation} 
%
Using usual properties of the flow of a vector field \cite{Book_Abraham_Marsden_Ratiu_TensorAnalysis}, for each $p \in \MS_{p_0}$, define $\I_p \subset \R$ the maximum open interval for which $\tau \to F_\normal^\tau (p)$ is defined, and $U=\cup_{\MS_{p_0}} \I_p \times \{p\}$, the maximal $\normal$-development of $\MS_{p}$, which is an open subset of $\M$. Since $F^{\tau}_\normal(p)$ is not a closed time function, each point $q \in U$ is uniquely defined by a pair $(\tau,p)$ with $p \in \MS_{p_0}$ and $\tau \in \I_p$. Therefore, the expression (\ref{Eq:ScaleFactorvsFluxNormal}) allows us to define the scale factor as a smooth function $a: U \to \R$ which is unique as we set $a_0=1$ in $\MS_{p_0}$.    

The rest of the proof follows by local analysis. Hence, take a connected and open neighbourhood $V \subset U$ around any $p \in U$ such that its intersections with the $\normal$-space sections are connected and simply connected. There the time function $t:V \to \R$ and the "potential" $\phi:V \to \R$ are well defined such that $\normal^\bot = \mathrm{Ker}(dt)$, $(\partial/\partial t)^\flat=dt$, $\normal = e ^{-\phi}\partial/\partial t$, and $\grad_\normal\normal=\gradbar\phi$. In this case, $\hubble = e ^{-\phi}\partial(\ln a)/\partial t$ and
\begin{equation}
\hubblevec = \gradbar\hubble + \hubble \gradbar\phi 
= e^{-\phi}\gradbar\left(e^\phi \hubble\right) = e^{-\phi}\gradbar\left(\frac{1}{a}\frac{\partial a}{\partial t}\right)  \, . 
\end{equation}
Therefore, $\hubblevec=0$ in $V$ if, and only if, there is a time function $f(t)$ such that $a=e^{f(t)}$ with $f=0$ in $\MS_{p_0}$. Moreover, for any volumetric scale factor $a:U \to \R$, $\widehat{\Omega}(\submetric) = a^{m-1} \widehat{\Omega}(\submetrict)$,  $\Lie_\normal\widehat{\Omega}(\submetric) = (m-1)\hubble \widehat{\Omega}(\submetric)= (m-1)\,\left(a^{-1}\Lie_\normal a\right) \, \widehat{\Omega}(\submetric)$, and
\begin{equation}
\Lie_\normal\widehat{\Omega}(\submetrict) = \frac{1}{a^{m-1}}\left(\Lie_\normal\widehat{\Omega}(\submetric) - (m-1)\left(\frac{1}{a}\Lie_\normal a\right) \widehat{\Omega}(\submetric) \right)
= 0  \, . 
\end{equation}
This finishes the proof of (i).  

With no loss of generality, we can assume $V$ to be the domain of adapted coordinates $(t,x^i)$ where the metric is represented as $\metric = - \, e^{2 \phi(t,x)}\, dt^2 +  \submetric_{ij}(t,x) \, dx^i dx^j$. In this case, the expansion tensor is 
\begin{equation}\label{Eq:DefExpansionTensor}
\theta_{ik} =\frac{e^{-\phi}}{2}\frac{\partial}{\partial t}\submetric_{i k} = \hubble\,\submetric_{i k} + \shear_{i k} \, .    
\end{equation}
The isotropy hypothesis of the expansion/contraction is expressed as 
\begin{equation}
\shear_{i k}=0
\quad \iff \quad 
\frac{\partial\submetric_{ij}}{\partial t} = 2 \, \frac{1}{a}\frac{\partial a}{\partial t} \, \submetric_{ij}
 \, .
\end{equation}
But that is equivalent to $\submetric_{ij}(t,x) = a(t,x)^2 \submetrict_{ij}(x)$. A simple calculation tells us that $\Lie_\normal\submetrict=\Lie_\normal\left(\submetrict_{ik}(t,x)dx^idx^k\right)=0$ if, and only if, $\partial \submetrict_{ik}/\partial t=0$, thus proving the part (ii). Part (iii) follows straightforwardly from (i) and (ii).
\qed

The volumetric scale parameter of theorem \ref{Thm:HubbleHomogeneityIsotropy} can be defined globally in $\M$ as far as we ensure the existence of a global cosmic time $t: \M \to \R$ such that the $\normal$-spatial sections are characterized by $t=\textrm{const}$, as in the globally hyperbolic case \cite{CMP_2005_GlobalHyperbolic}, for instance. In general, it cannot be extended to the whole spacetime, and even if that happens, the $\normal$-spatial sections may not be Cauchy hypersurfaces (see sec. 3 in Ref. \cite{GRG_2022_Sanchez}). 

From here on, let us assume our spacetime to admit a global splitting $\M = \I \times \MS$ following the existence of observers $\normal$ to whom the Hubble evolution is homogeneous and isotropic, with $t: \I \times \MS \to \I$, the canonical projection $(t,x) \mapsto t$, characterizing the $\normal$-spatial sections as $t=\textrm{const}$. According to the theorem \ref{Thm:HubbleHomogeneityIsotropy}, we have  
\begin{equation}\label{Eq:MetricGeneralScaleFactor}
\metric = - e^{2\phi}dt^2 + a(t)^2\, \submetrict \, , 
\end{equation}
with $\submetrict$ a Riemannian metric in $\MS$ and $\phi:\I \times \MS \to \R$ the function defined by $e^{2\phi}=-\metric (\grad t, \grad t)$, which satisfies $\gradbar\phi = \grad_\normal\normal$, and thus resembles the Newtonian gravitational potential \cite{LGGomes_2024_CQG_1}. The Hubble parameter takes the form  
\begin{equation}\label{Eq:HubbleStandardModel}
\hubble = e^{-\phi}\, \hubble_S
\qquad \textrm{with} \qquad \hubble_S = \frac{1}{a}\, \frac{da}{dt} \, .
\end{equation}
As we denote  Levi-Civita connection of $\submetrict$ by $\gradbarD$, the $\submetrict$-gradient $\gradbarD f$, the $\submetrict$-Laplace-Beltrami operator $\gradbarD^2 f$, and the mean sectional curvature $K$ of $\submetrict$ are given by, respectively,
\begin{equation}
\gradbarD f = a^2\, \gradbar f
\, , \quad
\gradbarD^2 f = a^2\, \gradbar^{\, 2} f \, ,
\quad \textrm{and} \quad
\curvature = (m-1)(m-2)\, \frac{K}{a^2}  \, ,
\end{equation}
where $\curvature$ is the scalar curvature of $\submetric=a^2\, \submetrict$. Einstein's equations, following the comoving approach \cite{Book_2012_ellis_mac_marteens}, begins with the generalized Friedmann equation, 
\begin{equation}\label{Eq:FriedmannEquation}
\frac{1}{2}(m-1)(m-2)\left( \hubble^2 + \frac{K}{a^2} \right) = 
\energy \, ,
\end{equation}
and the Raychaudhuri equation, written in terms of the ``potential" $\phi$,
\begin{equation}\label{Eq:RaychaudhuriPhi}
\frac{1}{m-1}\left(\gradbarD^2\phi + |\gradbarD \phi|^2\right) 
= \frac{a^{3-m}}{2\, \hubble_S} \frac{\partial}{\partial t}\left(e^{-2\phi}\,a^{m-1}\,\hubble^2_S \right)   
  + \frac{(m-3)}{2}K + \frac{\pressure\, a^2}{m-2} \, . 
\end{equation}
Here, $\energy = T_{\mu\nu}\normal^\mu\normal^\nu$ is the energy density, $\pressure = \left(T^\mu_\mu + \energy\right)/3$ the relativistic pressure, and $|\gradbarD \phi|^2=\submetrict_{ik}\gradbarD^i \phi\, \gradbarD^k \phi$. These two equations imply the conservation of energy, $\normal_\nu \grad_\mu T^{\mu\nu}=0$, just as in the FLRW model, which turns out to be
\begin{equation}\label{Eq:EnergyConservation}
\frac{\partial\energy}{\partial a} + \frac{m-1}{a}\left( \energy + \pressure \right) 
= \frac{(m-2)}{a^3}\, \left( \gradbarD^2\phi + |\gradbarD \phi|^2\right) \, .
\end{equation}
Note that in dimension $m=4$, in the limit of a nearly static evolution of the universe, $\hubble^2 << \energy$ and $a \approx 1$, the generalized Friedmann equation implies $3\, K \approx \energy$ and therefore, from (\ref{Eq:RaychaudhuriPhi}),
\begin{equation}\label{Eq:NewtonianLimit}
\gradbarD^2\phi + |\gradbarD \phi|^2  
\approx \frac{\energy+3\pressure}{2}  \, . 
\end{equation}
As we fix $p \in \MS$ and consider only the points of $\MS$ at small $\submetrict$-distances compared to the length $K(p)^{-1/2}$, we can use normal coordinates to obtain $\submetrict_{ik} \approx \delta_{ik}$ \cite{MTW}. Therefore, the weak field limit $\phi <<1$ along with the condition of non-relativistic pressure $\pressure << \energy$ give us the Newtonian limit near $p$ at distances $d << K(p)^{-1/2}$.

The energy flux equation, $``G_0^i = T_0^i$", is
\begin{equation}\label{Eq:EnergyFluxEquation}
q^i = \frac{(m-2)}{a^2}\,  \gradbarD^i \hubble  \, .
\end{equation}
It tells us that, in the case of $\hubble >0$, energy must propagate in the direction where the expansion is optimal. If we appeal to the Newtonian interpretation of $\phi$, we conclude that the energy flux points to the direction of the "gravitational force" $\gradbarD\phi$. In particular, there is no convection of energy along the "equipotential" hypersurfaces $\phi=\textrm{const}$. 

In the spirit of interpreting $\phi$ as the gravitational potential of the nearly Newtonian regime $\phi<<1$, we recover the trace-free part of $\gradbar_i\gradbar_{k}\phi$, as expected for the Newtonian tidal tensor \cite{Book_2013_Ohanian}, by 
\begin{equation}\label{Eq:TidalTensor}
\widehat{\Phi}_{ik}:= e^{-\phi}\, \gradbar_i\gradbar_{k}e^{\phi} - \frac{1}{m-1}\, e^{-\phi}\, \gradbar^2 e^{\phi} \, \submetric_{ik}  \, ,
\end{equation}
where $\widehat{\Phi}_k^k=0$. In terms of it, the remaining Einstein's equations are the trace-fre part of the spatial components, which we read as 
\begin{equation}\label{Eq:HubbleAnisotropyEquation}
\Pi_{ik} + \widehat{\Phi}_{ik} = \widehat{R}_{ik} \, ,
\end{equation}
where $\widehat{R}_{ik}$ is the trace-free part of the Ricci tensor of the metric $\submetrict$. Whether the spatial geometry is Einstein, that is, $\widehat{R}_{ik}=0$, the equation (\ref{Eq:HubbleAnisotropyEquation}) would be interpreted as an equilibrium relation between the inner tensions in matter and the Newtonian-like tidal forces acting upon it. In general, it tells us that the net "residual forces" obtained by the excess of the gravitational "tidal forces" over the inner stresses they cause upon the matter content are accumulated in the form of space curvature.

We claim that the class of spacetimes admitting homogeneous and isotropic expansion is large enough and physically meaningful to allow interesting applications to cosmology. In particular, if $\submetrict$ has constant curvature, we can follow the ideas presented in Ref. \cite{LGGomes_2024_CQG_1}, and choose suitable periodic boundary conditions such that our spacetime becomes homogeneous and isotropic on large scales \cite{CliftonFerreira_2009, Hellaby_2012, Bruneton_2012, Zhuk_2015, Eingorn2021, LGGomes_2022_CQG_2}. In this case, an effective FLRW-like cosmological model appears after we take average values along those extensions delimited by the homogeneity scale $L_0$, the so-called cosmological cells, thus leading to new terms arising from local inhomogeneities. All these ideas lead us to new proposals in theoretical cosmology. By the time of writing this manuscript, they are under investigation and will soon appear in the literature \cite{LGGomes_2024_2}.     

The author is thankful for the support from FAPEMIG, project number RED-00133-21. \vspace{1cm}

\bibliography{ref_IHIS.bib}

\end{document}